\documentclass[
  aps,%
  pra,%
  reprint,%
]{revtex4-2}

\usepackage{amsmath}
\usepackage{amssymb}
\usepackage{amsthm}
\usepackage{bm,bbm}

\usepackage{physics}

\usepackage[T1]{fontenc}
\usepackage[sc,osf]{mathpazo}

\usepackage{xcolor}
\usepackage{graphicx}

\usepackage[bookmarks=false]{hyperref}
\hypersetup{colorlinks=true,citecolor=blue,linkcolor=red,%
urlcolor=blue,pdfstartview=FitH,bookmarksopen=true}

\DeclareMathOperator{\diag}{diag}

\newcommand{\vect}[1]{\bm{#1}}
\newcommand{\ee}{\mathrm{e}}
\newcommand{\ii}{\mathrm{i}}

\theoremstyle{definition}

\newtheorem{obs}{Observation}

\newcommand{\er}{\varepsilon}
\newcommand{\sx}{\sigma_x}
\newcommand{\sy}{\sigma_y}
\newcommand{\sz}{\sigma_z}
\newcommand{\mi}{\mathbb{I}}
\newcommand{\ot}{\otimes}

\newcommand{\ope}[3]{{#1}_{#2}^{\qty(#3)}}

\begin{document}

\title{Protecting entanglement witnesses with randomized measurements}

\author{Jing-Tao Qiu}
\affiliation{Department of Physics, Shandong University, Jinan 250100, China}

\author{Wei-Jie Jiang}
\affiliation{Department of Physics, Shandong University, Jinan 250100, China}

\author{Xiao-Dong Yu}
\email{yuxiaodong@sdu.edu.cn}
\affiliation{Department of Physics, Shandong University, Jinan 250100, China}

\date{\today}

\begin{abstract}
  Entanglement is one of the most prominent features of quantum mechanics and
  serves as an essential resource in quantum information science. Therefore,
  the certification of entanglement is crucial for quantum information
  processing tasks. While entanglement witnesses are the most frequently used
  method for entanglement certification in experiments,
  recent research shows that even tiny errors in
  measurements may significantly undermine the effectiveness of a witness.
  In this work, we propose a randomized-measurement-based method
  to solve this problem. Through this method, the errors
  in measurement results can be substantially suppressed,
  thereby restoring the certification capability of entanglement witnesses.
  Our method is not only applicable to general types of witnesses, including
  multi-party entanglement and high-dimensional entanglement witnesses, but
  also experimentally friendly in the sense that only slight modifications are
  needed to the original measurement settings.
\end{abstract}

\maketitle

\section{Introduction}
Entanglement, as one of the most crucial resources in quantum information
processing \cite{HorodeckiQuantumentanglement2009}, holds great importance in
many fields, including quantum cryptography
\cite{Scaranisecuritypracticalquantum2009,XuSecurequantumkey2020,
pirandolaAdvancesQuantumCryptography2020},
quantum computation
\cite{Jozsaroleentanglementquantum2003,BriegelMeasurementbasedquantum2009},
and quantum sensing
\cite{GiovannettiAdvancesquantummetrology2011,DegenQuantumsensing2017}.
Therefore, certifying the presence of entanglement is a central topic in
quantum information \cite{GuehneEntanglementdetection2009,
FriisEntanglementcertificationtheory2019}.
Among various certification methods,
entanglement witnesses are the most commonly used one,
since they can be directly measured in experiments and often with only few
measurement settings
\cite{GuehneEntanglementdetection2009,FriisEntanglementcertificationtheory2019}.

State-of-the-art technology guarantees highly accurate operations on quantum
systems. Nevertheless, errors are inevitable in all processes, which can
undermine the effectiveness of quantum information processing tasks.
For entanglement witnesses, it has been shown that even tiny errors in
measurements can already cause significant deviation from the precise result,
thereby undermining the effectiveness of the witnesses
\cite{SeevinckLocalcommutativityversus2007,RossetImperfectmeasurementsettings2012}.
One way to circumvent this problem is the so-called device-independent method
\cite{BancalDeviceIndependentWitnesses2011,MoroderDeviceIndependentEntanglement2013}.
In this method, one can certify the presence of entanglement without relying on
the detailed knowledge of the measurement apparatus.
However, this method comes with not only the theoretical drawback that many
entangled states are in principle not certifiable, but also
the practical drawback that experiments become much more challenging.
In fact, the device-independent method overestimates the errors in measurements,
disregarding the prior knowledge on the actual measurements.
Alternatively, Morelli \textit{et al.} quantitatively investigated the relation
between the infidelity of measurements and the minimum expected value of
a given entanglement witness for separable states
\cite{MorelliEntanglementDetectionImprecise2022,CaoGenuineMultipartiteEntanglement2024}.
With this relation, one can certify the presence of entanglement if the witness
result is smaller than this minimum value. While this method takes full
advantage of the prior knowledge of the measurement infidelity, it sacrifices
a large portion of the certification capability, and even completely invalidates
the witness for some small infidelities.

In recent years, randomized measurements have been adopted more and more
frequently in diverse quantum information processing tasks
\cite{Elbenrandomizedmeasurementtoolbox2023, CieslinskiAnalysingquantumsystems2024}, such as characterization of
topological order \cite{ElbenManybodytopological2020} 
, machine
learning for quantum many-body problems
\cite{HuangProvablyefficientmachine2022}, and fidelity
estimation of quantum states or quantum circuits
\cite{HuangPredictingmanyproperties2020,
SilvaPracticalCharacterizationQuantum2011,
YuStatisticalMethodsQuantum2022}.
This comes from the distinct advantages of randomized measurements,
including simplifying the noise model
\cite{WallmanNoisetailoringscalable2016,HashimRandomizedCompilingScalable2021,
ChenRobustShadowEstimation2021,KohClassicalShadowsNoise2022}
and reducing of the resource consumption
\cite{EmersonScalablenoiseestimation2005,
KnillRandomizedbenchmarkingquantum2008,
MagesanScalableRobustRandomized2011}.

In this work, we find that randomized measurements can also be used to
suppress errors in the measurement result. Based on this, we propose a method to
suppress errors in entanglement witnesses.
Our method can restore most of the certification capability, and significantly
enhance the robustness of witnesses to measurement errors.
Moreover, the method is experimentally friendly, since it does not cost extra
copies of the quantum states and compared to the original measurement settings
only some local unitary operations are supplemented.

\section{Entanglement witnesses with imprecise measurements}
Entanglement witnesses are a fundamental tool in entanglement theory
\cite{HorodeckiSeparabilitymixedstates1996,
TerhalBellinequalitiesseparability2000}.
They are observables that completely characterize the set of entangled states, in the sense that any entangled state can be certified by some witness.
Mathematically, an entanglement witness $W$ is an observable satisfying that
$\Tr(W\rho_s) \ge 0$ for all separable states $\rho_s$.
Thus, if the expected value $\Tr(W\rho)$ obtained from experiments is
negative, we directly certify that $\rho$ is an entangled state.
In actual experiments, the measurements are, however, never perfect.  These
errors will undermine the certification capability or even the effectiveness of
entanglement witnesses.

To quantitatively investigate the impact of measurement errors, we employ the
notion of measurement infidelity \cite{StanoReviewperformancemetrics2022}.
Suppose that we aim to perform a projective measurement
$\qty{P_i}_{i=1}^d$ (called the target measurement) on a $d$-dimensional
quantum system, but inevitable errors will always make the measurement
imprecise.  This imprecise measurement shall be described by
a positive-operator-valued measure (POVM) $\qty{M_i}_{i=1}^d$
(called the laboratory measurement). The difference between the target measurement and
the laboratory measurement can be quantified by the
measurement infidelity
\begin{equation}
	\er=1-\frac{1}{d}\sum_{i=1}^d\Tr(P_iM_i).
	\label{eq:inaccuracy}
\end{equation}
From the definition, one can easily see that $0\le\er\le1$, and $\er=0$ if and
only if the laboratory measurement is identical to the target
measurement. An important advantage of measurement infidelity is that this
quantity can be directly measured in experiments
\cite{MorelliEntanglementDetectionImprecise2022}.

When the measurements are imprecise, so will the measurement result of the
witness $W$. As a result, there may exist separable states such that the
measurement results are negative, making the entanglement certification
unreliable. One may think that this kind of errors is negligible when the
measurement infidelities are small. This is however not true.
Previous works show that even tiny measurement infidelities can severely
compromise entanglement witnesses
\cite{SeevinckLocalcommutativityversus2007,RossetImperfectmeasurementsettings2012,
MorelliEntanglementDetectionImprecise2022}.
For example, the witness $W=\mi_4-\sx^A\otimes\sx^B-\sz^A\otimes\sz^B$ can
certify the presence of entanglement in a two-qubit system when the expected
value $\Tr(\rho W)$ is in the range $[-1,0)$.
However, a tiny measurement infidelity $\er=0.5\%$ can lead the minimum
expected value for separable states to be $-0.279$,
compromising over a quarter of the original certification range $[-1,0)$  \cite{MorelliEntanglementDetectionImprecise2022}.

\section{Error suppression with randomized measurements}
To obtain an efficient error suppression method, we first show that there
exists an asymmetric structure in measurement errors,
which serves as the cornerstone of our error suppression method.
In order to present the idea, we consider the target measurement $\sz$ on
a qubit system. The generalization to general projective measurements will be
provided in the next section.
The target measurement $\sz$ corresponds to the projective measurement
$\qty{\ketbra{0},\ketbra{1}}$ with outcomes $\pm 1$.
In actual experiments, the imprecise laboratory measurement shall be
described by a general POVM $\qty{M_+,M_-}$, which always admits the
decomposition
\begin{align}
    M_\pm&=\frac{1\pm p}{2}\mi_2 \pm \frac{n_x\sx+n_y\sy+n_z\sz}{2},
\end{align}
where $p,n_x,n_y,n_z$ are real numbers, and $\mi_2,\sx,\sy,\sz$ are the
identity and Pauli operators.
When the quantum system is in state $\rho$, the expected value is given by
$\Tr(M_+\rho)-\Tr(M_-\rho)=\Tr(M\rho)$, where
\begin{equation}
  M=M_+-M_-=p\mi_2+n_x\sx+n_y\sy+n_z\sz.
  \label{eq:M}
\end{equation}
By Eq.~\eqref{eq:inaccuracy}, the infidelity of the measurement infidelity
equals to
\begin{equation}
	\er=1-\frac{\expval{M_+}{0}+\expval{M_-}{1}}{2}=\frac{1-n_z}{2}.
	\label{eq:inaccuracyQubit}
\end{equation}
Let us write $\qty(n_x,n_y,n_z)$ as a vector $\vect{n}$,  then the positivity
constraint $M_+\ge 0$ and $M_-\ge 0$ is equivalent to the condition
\begin{equation}
  |p|+\norm{\vect{n}}\le1,
  \label{eq:constrantPN}
\end{equation}
where $\norm{\vect{n}}=\sqrt{n_x^2+n_y^2+n_z^2}$ is
the Euclidean norm. Thus, Eqs.~(\ref{eq:inaccuracyQubit},\,\ref{eq:constrantPN})
impose a constraint on the range of $p$ and $\vect{n}$.
Specifically, we have that
$n_x,n_y\in\qty[-2\sqrt{\er\qty(1-\er)},2\sqrt{\er\qty(1-\er)}]$ and $p\in
\qty[-2\er,2\er]$.

Note that the deviation of the laboratory measurement from the target measurement is
given by
\begin{equation}
  M-\sz=p\mi_2+n_x\sx+n_y\sy+\qty(n_z-1)\sz.
  \label{eq:deviation}
\end{equation}
The discussion above shows that the magnitudes of noncommutative terms
$n_x\sx,n_y\sy$ and those of commutative terms $p\mi,(n_z-1)\sz$ in
Eq.~\eqref{eq:deviation} behave quite differently. Here and in the following,
the commutativity is always with respect to the corresponding target
measurement. Particularly, in the regime of small
infidelities (i.e., $\er\rightarrow0$), the largest magnitudes of
$n_x,n_y$ scale as $\sqrt{\er}$, while those of $p$ and $n_z-1$ only scale as
$\er$.  Thus, the noncommutative terms are dominant for the laboratory measurement
errors. Indeed, it is exactly this square root dependence that results in tiny
measurement infidelities severely compromising entanglement witnesses.

In the following, we will show that the noncommutative terms can be eliminated
via randomized measurements, thereby significantly suppressing the errors in
entanglement witnesses.
The key to our randomized-measurement-based method is that
before performing the laboratory measurement $M$, a randomized phase gate
$U_\theta=\diag(1,\ee^{\ii\theta})$ under the measurement basis
is performed, where the random
variable $\theta$ is chosen uniformly from $[0,2\pi)$.
This is equivalent to performing a randomized measurement
$U_\theta^\dagger MU_\theta$ on the quantum system, as
$\Tr(U_\theta\rho U_\theta^\dagger M)=\Tr(\rho U_\theta^\dagger MU_\theta)$.
Furthermore, the expected value of this randomized measurement on a quantum
state $\rho$ is $\frac{1}{2\pi}\int_0^{2\pi}\Tr(\rho U_\theta^\dagger
MU_\theta)\dd{\theta} =\Tr(\rho\bar{M})$
where
\begin{equation}
  \bar{M} = \frac{1}{2\pi}\int_0^{2\pi}U_\theta^\dagger
  MU_\theta\dd{\theta},
  \label{eq:defMbar}
\end{equation}
which we will refer to as the tuned measurement.

Let us analyze how the randomization procedure affects the laboratory measurement $M$
in Eq.~\eqref{eq:M}. Note that for the target measurement $\sz$, the
measurement basis is $\{\ket{0},\ket{1}\}$ and thus
$U_\theta=\ketbra{0}+\ee^{\ii\theta}\ketbra{1}$.
On the one hand, as $\mi_2$ and $\sz$ are invariant
under $U_\theta^\dagger$, i.e., $U_\theta^\dagger \mi_2U_\theta=\mi_2$ and
$U_\theta^\dagger \sz U_\theta=\sz$, the commutative terms $p\mi_2$ and
$n_z\sz$ remain unchanged.
On the other hand, as $\int_0^{2\pi} U_\theta^\dagger \sx
U_\theta\dd{\theta}=\int_0^{2\pi}(\sx\cos\theta-\sy\sin\theta)\dd{\theta}=0$ and
$\int_0^{2\pi} U_\theta^\dagger \sy
U_\theta\dd{\theta}=\int_0^{2\pi} (\sx\sin\theta+\sy\cos\theta) \dd{\theta}
=0$, the noncommutative terms $n_x\sx$ and $n_y\sy$ are eliminated.
Thus, the randomization procedure can achieve the goal of eliminating the
noncommutative terms, i.e., $\bar{M} =  p\mi_2 + n_z\sz$,
thereby eliminating the dominant errors in the lab
measurement.

\begin{figure}
  \includegraphics[width=.8\linewidth]{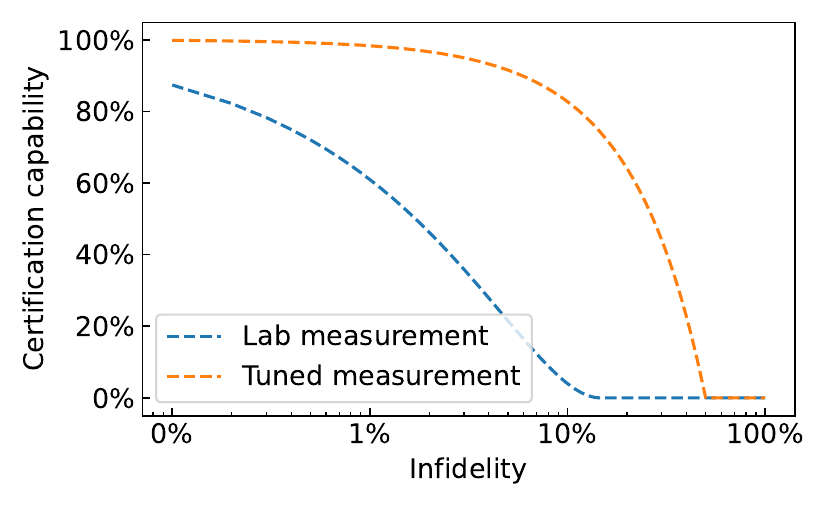}
  \caption{The certification capability
    of the witness $W=\mi_4-\sx^A\otimes\sx^B-\sz^A\otimes\sz^B$ under
    different infidelities.
    }
  \label{fig:comparison}
\end{figure}

To show the advantage of our method, we consider the two-qubit entanglement
witness $W=\mi_4-\sx^A\otimes\sx^B-\sz^A\otimes\sz^B$. In this
case, the target measurements are $\sx$ and $\sz$ on each qubit. When the
measurements are perfect, the witness $W$ can certify
the presence of entanglement if the expected value is in the range $[-1,0)$,
which we call the certification range of $W$. Now, consider the case that the
measurements are imprecise. Quantitatively, we suppose that the
infidelities of all laboratory measurements are at most $\er$, then the minimum
expected value for separable states is
\cite{MorelliEntanglementDetectionImprecise2022}
\begin{equation}
    \mathcal{B}(\er) = -4(1-2\er)\sqrt{\er(1-\er)}
    \label{eq:ImpBound}
\end{equation}
when $\er\le(2-\sqrt{2})/4$ and $\mathcal{B}(\er)=-1$ otherwise.
For the nontrivial case that $\mathcal{B}(\er)>-1$
the minimization is achieved when the corresponding laboratory measurements
$M_x^A,M_z^A,M_x^B,M_z^B$ are
\begin{equation}
  M_{x/z}^A=M_{x/z}^B=(1-2\er)\sigma_{x/z}+2\sqrt{\er(1-\er)}\sigma_{z/x}.
  \label{eq:ImpSaturation}
\end{equation}
One can easily see that $\mathcal{B}(\er)\approx-4\sqrt{\er}$ in the regime of
small infidelity, which results from the noncommutative terms in
Eq.~\eqref{eq:ImpSaturation}.

While with the randomized measurements, the noncommutative terms are eliminated
and thus the tuned measurements are of the form
$\bar{M}_{x/z}=p_{x/z}\mi+\lambda_{x/z}
\sigma_{x/z}$, where the parameters $\lambda_x=\lambda_z=1-2\er$ and
$\abs{p_x}$ and $\abs{p_z}$ are no larger
than $2\er$ by Eqs.~(\ref{eq:inaccuracyQubit},\,\ref{eq:constrantPN}).
In fact, the minimum expected value for separable states can be achieved when
$p_x=p_z=2\er$ (given that $\er\le 1/2$), i.e., the tuned measurements
$\bar{M}_x^A,\bar{M}_z^A,\bar{M}_x^B,\bar{M}_z^B$
are
\begin{equation}
  \bar{M}_{x/z}^A=\bar{M}_{x/z}^B=2\er\mi_2+(1-2\er)\sigma_{x/z},
  \label{eq:ModifiedSaturation}
\end{equation}
and the minimum value is
\begin{equation}
  \mathcal{B}^{\mathrm{rand}}(\er)
  =-4\qty(\sqrt{2}-1)\er-4\qty(3-2\sqrt{2})\er^2.
  \label{eq:ModifiedBound}
\end{equation}
When $\er>1/2$, $\mathcal{B}^{\mathrm{rand}}(\er)=-1$.
See Appendix~\ref{appendix:bi} for the proof.
Obviously, the randomized measurements result in a much better bound
$\mathcal{B}^{\mathrm{rand}}(\er)\approx-4\qty(\sqrt{2}-1)\er$ in the regime of
small infidelity.

When imprecise laboratory measurements are preformed and the
obtained result of the witness is in the range $[\mathcal{B}(\er),0)$,
chances are that the quantum system is in a separable state. As a result,
the remaining certification range turns to $[-1,\mathcal{B}(\er))$. Similarly,
for the tuned measurements, the remaining certification range turns to
$[-1,\mathcal{B}^{\mathrm{rand}}(\er))$. The different scaling of
$\mathcal{B}(\er)$ and $\mathcal{B}^{\mathrm{rand}}(\er)$ in
Eq.~\eqref{eq:ImpBound} and Eq.~\eqref{eq:ModifiedBound} indicates that our
randomized-measurement-based method can significantly suppress the errors in
entanglement witnesses.
In Fig.~\ref{fig:comparison}, we show the detailed comparison of
the results before and after applying our method, where we have used the ratio
of the remaining certification range
to the original to quantify the certification capability.
As a concrete example, when $\er=0.5\%$,  the laboratory measurement results
in $\mathcal{B}=-0.279$, while our method can significantly reduce the error to
$\mathcal{B}^\mathrm{rand}=-0.008$, compromising $27.9\%$ and $0.8\%$ of the
certification range, respectively.

\section{General results}
The square root deviation of the measurement results to the measurement
infidelity is not limited to the qubit case, but quite common for the actual
measurements of any dimension. For example, the square root dependence always
presents for the misaligned measurements. This kind of errors emerges due to
the loss of perfect control and is widely encountered in actual experiments
\cite{RossetImperfectmeasurementsettings2012,
CaoGenuineMultipartiteEntanglement2024}.
Quantitatively, we have the following observation:
  Let $M=\sum_{i=1}^d\lambda_i\ketbra{\varphi_i}$ be a nondegenerate projective
  measurement. If the laboratory measurement is a misaligned measurement
  $\tilde{M}=\sum_{i=1}^d\lambda_i\ketbra{\tilde\varphi_i}$
  with infidelity $\er$,
  then $\norm{\tilde{M}-M}\ge\lambda\sqrt{2\er}$, where
  $\lambda=\min_{i\ne j}|\lambda_i-\lambda_j|$ and
  $\norm{\cdot}$ denotes the Hilbert-Schmidt norm.
See Appendix~\ref{appendix:misalignment} for the proof.

Fortunately, the randomized measurement provides a general
method of error suppression.
The key point is, for a general $d$-dimensional quantum system, we can still
eliminate all the noncommutative terms of any laboratory measurement with randomized
measurements.

Suppose that we aim to perform the target measurement
$\qty{P_i=\ketbra{\varphi_i}}_{i=1}^d$ on a $d$-dimensional quantum system with
corresponding outcomes $\qty{\lambda_i}_{i=1}^d$, where
$\qty{\ket{\varphi_i}}_{i=1}^d$ is an orthonormal basis.
In experiments, the laboratory measurement $\qty{M_i}_{i=1}^d$ is actually
performed due to the existence of measurement errors. To suppress the errors,
we perform a randomized unitary operation $U_{\vect{\theta}}=\sum_{k=1}^d
\ee^{\ii\theta_k} \ketbra{\varphi_k}$ before the laboratory measurement. The random
variables $\theta_1,\theta_2,\dots,\theta_d$ are chosen uniformly and
independently from $[0,2\pi)$. This is equivalent to performing the randomized
measurement $\qty{U_{\vect{\theta}}^\dagger M_i U_{\vect{\theta}}}_{i=1}^d$, as
$\Tr(U_{\vect{\theta}}\rho U_{\vect{\theta}}^\dagger M_i)=\Tr(\rho
U_{\vect{\theta}}^\dagger M_i U_{\vect{\theta}})$.  In this way, the expected
value of the measurement outcomes on a quantum state $\rho$ is
$\mathbb{E}_{\vect{\theta}}\qty[\sum_{i=1}^d \lambda_i \Tr(\rho
U_{\vect{\theta}}^\dagger M_i U_{\vect{\theta}})]=\sum_{i=1}^d \lambda_i
\Tr(\rho \bar{M}_i)$,
where $\mathbb{E}_{\vect{\theta}}$ denotes the expected value over the uniform
and independent choices of $\theta_1,\theta_2,\dots,\theta_d\in[0,2\pi)$, and
$\bar{M}_i=\mathbb{E}_{\vect{\theta}}\qty(U_{\vect\theta}^\dagger M_i
U_{\vect{\theta}})$ are the POVM elements of the tuned measurements.
As $\qty{\ket{\varphi_k}}_{k=1}^d$ is an orthonormal basis,
$U_{\vect{\theta}}^\dagger M_i U_{\vect{\theta}}$ admits the form $\sum_{k,\ell=1}^d
\mel{\varphi_k}{M_i}{\varphi_{\ell}}\ee^{-\mathrm{i}\qty(\theta_k-\theta_{\ell})}
\ketbra{\varphi_k}{\varphi_{\ell}}$. Note also that
$\mathbb{E}_{\vect{\theta}}\qty[\ee^{-\mathrm{i}
\qty(\theta_k-\theta_{\ell})}]=\delta_{k\ell}$,
therefore
\begin{equation}
  \bar{M}_i = \sum_{k=1}^d \expval{M_i}{\varphi_k} \ketbra{\varphi_k}.
  \label{eq:Mbar-d}
\end{equation}
Thus, $\bar{M}_i$ is diagonal in the basis $\qty{\ket{\varphi_k}}_{k=1}^d$ and
shares the same diagonal elements with $M_i$. Then, Eq.~\eqref{eq:inaccuracy}
implies that the infidelity of the tuned measurement $\qty{\bar{M}_i}_{i=1}^d$
is identical to the laboratory measurement $\qty{M_i}_{i=1}^d$.

With the tuned measurements, the errors in entanglement witnesses can always be
suppressed into the scale of at most $\er$. Suppose that we aim to perform
a witness $W$ to a $N$-partite quantum system, which is measured via local
projective measurements, then the witness always admits the decomposition
\cite{GuehneEntanglementdetection2009}
\begin{equation}
  W=\sum_\mu w_\mu \ope{P}{\mu}{1}\otimes \ope{P}{\mu}{2}\otimes\dots\otimes
  \ope{P}{\mu}{N},
  \label{eq:Wproj}
\end{equation}
where $w_\mu$ are real numbers, and each projector $\ope{P}{\mu}{n}$
corresponds to an outcome of some projective measurement on the $n$-th
subsystem. Due to the measurement errors, the observable actually measured is
\begin{equation}
  W_\er=\sum_\mu w_\mu \ope{M}{\mu}{1} \otimes \ope{M}{\mu}{2} \otimes \dots
  \otimes \ope{M}{\mu}{N},
  \label{eq:Wlab}
\end{equation}
where $\ope{M}{\mu}{n}$ is the POVM element of the lab
measurement corresponding to $\ope{P}{\mu}{n}$ and the subscript $\er$ is used
to indicate that the infidelity of any laboratory measurement is at most $\er$. With
randomized measurements, the observable that we measure will turn to
\begin{equation}
  \bar W_\er=\sum_\mu w_\mu \ope{\bar{M}}{\mu}{1}\otimes \ope{\bar{M}}{\mu}{2}
  \otimes \dots \otimes \ope{\bar{M}}{\mu}{N},
  \label{eq:Wtuned}
\end{equation}
where $\ope{\bar{M}}{\mu}{n}$ are the POVM elements of the tuned measurement
corresponding to $\ope{M}{\mu}{n}$. Still, we define
\begin{equation}
  \mathcal{B}^{\mathrm{rand}}(\er):=\min_{W_\er}
  \min_{\rho\in\mathcal{S}}\Tr(\bar W_\er\rho),
  \label{eq:defBrand}
\end{equation}
where the first minimization is over $W_\er$ in Eq.~\eqref{eq:Wlab}, i.e.,
under the constraint that the infidelities of all measurements are no larger
than $\er$, and $\mathcal{S}$ is the set of separable states, or more
generally, the set of resource-free states. For example, for the certification
of multipartite entanglement, $\mathcal{S}$ denotes fully separable states; for
the certification of genuine multipartite entanglement, $\mathcal{S}$ denotes
biseparable states; and for the certification of $\ell$-dimensional
entanglement, $\mathcal{S}$ denotes all states with Schmidt number smaller than
$\ell$. In any case, the definition of witness implies that
$\mathcal{S}\subset\qty{\rho\mid\Tr(W\rho)\ge0}$, thus
$\mathcal{B}^{\mathrm{rand}}(\er)$ is bounded by
$\mathcal{B}^{\mathrm{rand}}(\er)\ge\min_{W_\er}
\min_{\Tr(W\rho)\ge0}\Tr(\bar W_\er\rho)$.
Furthermore, one can easily show that the difference between the
target measurement $\qty{P_i}_{i=1}^d$ and the tuned
measurement is $\qty{\bar{M}_i}_{i=1}^d$ is of order $\er$, more precisely,
\begin{equation}
  \qty(1-d\er)P_i\le\bar{M}_i\le\qty(1-d\er)P_i+d\er\mi_d,
  \label{eq:boundGeneralMbar}
\end{equation}
which further implies that $\mathcal{B}^{\mathrm{rand}}\qty(\er)$ is also of
order $\er$,
\begin{equation}
  \mathcal{B}^{\mathrm{rand}}\qty(\er)\ge\er\sum_{w_\mu<0}w_\mu\sum_{n=1}^N
  d_n+O\qty(\er^2),
  \label{eq:generalbound}
\end{equation}
where $d_n$ denotes the dimension of the $n$-th subsystem, and $O\qty(\er^2)$
are the higher-order errors. 
For the proofs of
Eqs.~(\ref{eq:boundGeneralMbar},\,\ref{eq:generalbound}), examples of
the high-dimensional system, and implementations with 
other sets of randomized measurements, please see Appendix~\ref{appendix:obs}.

\section{Gate errors and sampling errors}
In practice, the situation can be more complicated. On the one hand, the
randomization procedure can also be imprecise, which may bring extra errors in
the measurement results. This is relevant to gate errors. On the other hand,
one may wonder whether the randomized measurements consume more copies to
attain the desired precision, that is, the sampling errors of our method.
Analysis shows that our method is still effective under the influence of these errors.

Consider the gate errors first. Typically, the precision of the quantum
gate is one or two orders of magnitude better than the measurements in
the same physical system, such as in the superconducting systems
\cite{BaoFluxoniumAlternativeQubit2022,
  SomoroffMillisecondCoherenceSuperconducting2023,
  LiErrorpersingle2023,
  ZhouElectronchargequbit2024,
  WalterRapidHighFidelity2017,
ChenTransmonqubitreadout2023}
and semiconductor systems
\cite{StanoReviewperformancemetrics2022,
  Yonedaquantumdotspin2018,
  LawrieSimultaneoussinglequbit2023,
  BurkardSemiconductorspinqubits2023,
  WatsonAtomicallyengineeredelectron2017,
  HarveyCollardHighFidelitySingle2018,
  BlumoffFastHighFidelity2022,
TakedaRapidsingleshot2024}.
In this case, the impact of the gate errors on our randomized-measurement-based
method should be negligible.
This is confirmed by both theoretical and numerical analysis. 
See Appendices~\ref{appendix:gate_independent} and \ref{appendix:gate_dependent} for the details.
Moreover, the analysis demonstrates that our method
remains effective even when the precision of the quantum gates is comparable to
that of the measurements.

\begin{figure}
  \centering
  \includegraphics[width=.8\linewidth]{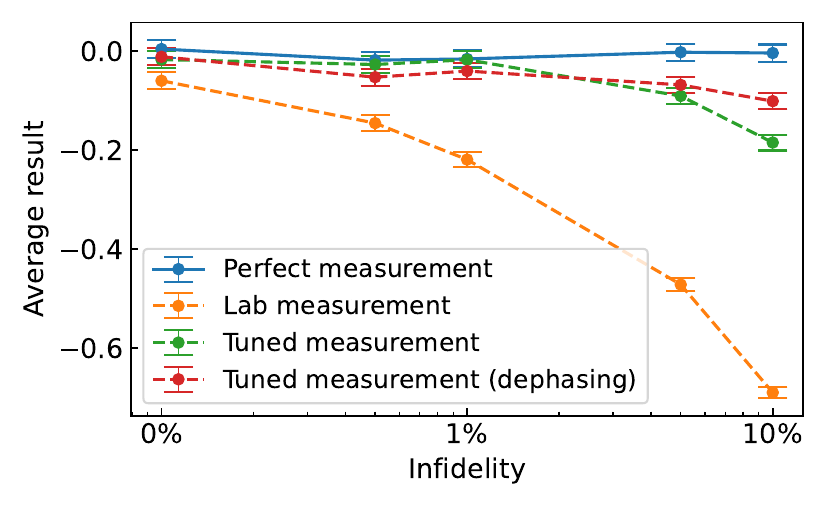}
  \caption{The simulation of results for the witness
    $W=\mi_4-\sx^A\otimes\sx^B-\sz^A\otimes\sz^B$.
    We have chosen five different values of the measurement infidelity $0.1\%$,
    $0.5\%$, $1\%$, $5\%$, and $10\%$. The dashed lines represent the worst
    cases for laboratory measurements and tuned measurements, while the solid line
    provides the results of perfect measurements as a reference. The error bar
    denotes the estimation of the standard deviation of the data.
  }
  \label{fig:sampling}
\end{figure}

When it comes to the sampling errors, one can anticipate that there will be no
obvious difference between the sampling errors of laboratory measurements and
randomized measurements due to their similar roles in the statistical analysis with Hoeffding's inequality \cite{HoeffdingProbabilityInequalitiesSums1963}.
This is further confirmed by the error bars in Fig.~\ref{fig:sampling}, where
each line is a simulation result with $5,000$ copies of the state. In the
figure, we also show the simulation result when the randomization procedure is
affected by dephasing errors, which further demonstrates the robustness of our
method to gate errors.

\section{Conclusion}
We proposed a method based on randomized measurements to suppress the
errors in entanglement witnesses caused by imprecise measurements.
Our method solves the problem that tiny infidelities in measurements
severely compromise the entanglement witnesses.
Through this method, the measurement errors can be substantially suppressed,
thereby restoring the certification capability of the witness.

Moreover, this method also exhibits both experimental friendliness and broad
applicability. On the one hand, one only needs to perform some local
unitary operations prior to the original measurement settings without
sacrificing the statistical significance of the original entanglement witness.
On the other hand, the method is applicable to general witnesses, including
those detecting multi-party entanglement and high-dimensional entanglement.

For future research, since the proposed randomized-measurement-based method
does not depend on the entanglement witnesses but only on the target
measurements, our method is expected to suppress errors in situations beyond
entanglement theory. 
For instance, it can help to improve the measurement error
mitigation, which is a widely used method for suppressing the errors in quantum
measurements, but it works exclusively for classical noises
\cite{MaciejewskiMitigationreadoutnoise2020}.
Assisted by randomized measurements, the influence of any noise can be reduced
to the form of classical noises such that measurement error mitigation can be
implemented \cite{CaiQuantumerrormitigation2023}.

\begin{acknowledgments}
  This work was supported by
  the National Natural Science Foundation of China
  (Grants No. 12205170 and No. 12174224)
  and the Shandong Provincial Natural Science Foundation of China
  (Grant No. ZR2022QA084).
\end{acknowledgments}

\section*{Data availability}
The data that support the findings of this work are openly available \cite{Qiujingtao1998Randomizedwitness}.

\onecolumngrid
\appendix

\section{$\mathcal{B}^{\mathrm{rand}}$ of the witness
$W=\mi_4-\sx^A\otimes\sx^B-\sz^A\otimes\sz^B$}\label{appendix:bi}

Since separable states are convex combinations of product states, the minimum
value can always be obtained on a product state $\rho=\rho^A\otimes\rho^B$.
That is,
\begin{equation}
  \mathcal{B}^{\mathrm{rand}}(\er)=\min_{\bar{M}_{x/z}^{A/B},\,\rho^{A/B}}
  \Tr[\qty(\mi_4-\bar{M}_x^A \otimes \bar{M}_x^B -\bar{M}_z^A\otimes
  \bar{M}_z^B)\rho^A\otimes \rho^B],
\end{equation}
where $\bar{M}_x^A$, $\bar{M}_z^A$, $\bar{M}_x^B$, and $\bar{M}_z^B$ are the
tuned measurements corresponding to the target measurements $\sigma_x^A$,
$\sigma_z^A$, $\sigma_x^B$, and $\sigma_z^B$, respectively, and the measurement
infidelities are at most $\er$. Let us denote $\Tr(\bar{M}_{x}^A\rho^A)$,
$\Tr(\bar{M}_{z}^A\rho^A)$, $\Tr(\bar{M}_{x}^B\rho^B)$, and
$\Tr(\bar{M}_{z}^B\rho^B)$ by $\expval{\bar{M}_{x}^A}$,
$\expval{\bar{M}_{z}^A}$, $\expval{\bar{M}_{x}^B}$, and
$\expval{\bar{M}_{z}^B}$, respectively, then we have
\begin{equation}
  \mathcal{B}^{\mathrm{rand}}(\er)=1-\max\qty(\expval{\bar{M}_x^A}
    \expval{\bar{M}_x^B}+\expval{\bar{M}_z^A} \expval{\bar{M}_z^B}),
\end{equation}
where the maximization is over $\expval{\bar{M}_x^A}$, $ \expval{\bar{M}_z^A}$,
$\expval{\bar{M}_x^B}$, $\expval{\bar{M}_z^B}$. By the Cauchy-Schwarz inequality,
we have $\expval{\bar{M}_x^A} \expval{\bar{M}_x^B}+\expval{\bar{M}_z^A}
\expval{\bar{M}_z^B}\le \sqrt{\expval{\bar{M}_x^A}^2
+\expval{\bar{M}_z^A}^2}\sqrt{\expval{\bar{M}_x^B}^2
+\expval{\bar{M}_z^B}^2}$.
Note that
\begin{equation}
  \max_{\expval{\bar{M}_x^A},\expval{\bar{M}_z^A}}
  \qty(\expval{\bar{M}_x^A}^2+\expval{\bar{M}_z^A}^2)
  =\max_{\expval{\bar{M}_x^B},\expval{\bar{M}_z^B}}
  \qty(\expval{\bar{M}_x^B}^2+\expval{\bar{M}_z^B}^2).
\end{equation}
Thus, we obtain that
\begin{equation}
  \mathcal{B}^{\mathrm{rand}}(\er)\ge 1-\max_{\expval{\bar{M}_x^A},
  \expval{\bar{M}_z^A}}\qty(\expval{\bar{M}_x^A}^2+\expval{\bar{M}_z^A}^2).
\end{equation}
The quantum state $\rho^A$ can be  written as
\begin{equation}
  \rho^A=\frac{\mi_2+r_x\sx+r_y\sy+r_z\sz}{2},
\end{equation}
where real numbers $r_x,r_y,r_z$ are components of the Bloch vector
$\vect{r}=(r_x,r_y,r_z)$ such that $\norm{\vect{r}}\le1$.
The tuned measurements are
\begin{equation}
  \bar{M}_x^A=p_x\mi_2+\lambda_x\sx,\quad\bar{M}_z^A=p_z\mi_2+\lambda_z\sz,
\end{equation}
where $|p_x|+|\lambda_x|\le1$, $|p_z|+|\lambda_z|\le1$ and
$\lambda_x,\lambda_z\in[1-2\er,1]$ by Eq.~\eqref{eq:constrantPN}.
Under these constraints, we also have $p_x,p_z\in[-2\er,2\er]$ when
$\er\le1/2$, then it holds that
\begin{equation}
  \begin{aligned}
    \mathcal{B}^{\mathrm{rand}}(\er)&\ge
    1-\qty(p_x + r_x \lambda_x)^2-\qty(p_z + r_z \lambda_z)^2\\
    &\ge 1-\qty(|p_x|+|r_x||\lambda_x|)^2-\qty(|p_z|+|r_z||\lambda_z|)^2\\
    &\ge 1-\qty[|r_x|+|p_x|(1-|r_x|)]^2-\qty[|r_z|+|p_z|(1-|r_z|)]^2\\
    &\ge 1-\qty[|r_x|+2\er(1-|r_x|)]^2-\qty[|r_z|+2\er(1-|r_z|)]^2.
  \end{aligned}
  \label{eq:BrandBoundA}
\end{equation}
Noting that the condition $\norm{\vect{r}}\le1$ implies that
$\abs{r_x}^2+\abs{r_z}^2\le 1$, one can easily show that
$\mathcal{B}^{\mathrm{rand}}\ge -4\er\qty[(\sqrt{2}-1)+\qty(3-2\sqrt{2})\er]$ from
the last line of Eq.~\eqref{eq:BrandBoundA}. Furthermore, one can directly
verify that $\mathcal{B}^{\mathrm{rand}}$ attains this value when
$r_x=r_z=\sqrt{2}/2$, $p_x=p_z=2\er$, and $\lambda_x=\lambda_z=1-2\er$.

For the case $\er>1/2$, the minimum value $\mathcal{B}^{\mathrm{rand}}(\er)=-1$
can always be obtained when
$\bar{M}_{x}^A=\bar{M}_{z}^A=\bar{M}_{x}^B=\bar{M}_{z}^B=\mi_2/2$.

\section{Misalignment errors and square root deviation}
\label{appendix:misalignment}

Misalignment of the measurement bases are inevitable errors in quantum
information processing due to the loss of perfect
control \cite{RossetImperfectmeasurementsettings2012,CaoGenuineMultipartiteEntanglement2024}.
In this appendix, we prove that misalignment always results in a square
root deviation; more precisely, we have the following observation.

\begin{obs}
  Let $M=\sum_{i=1}^d\lambda_i\ketbra{\varphi_i}$ be a nondegenerate projective
  measurement. If the laboratory measurement is a misaligned measurement with
  infidelity $\er$, i.e., $\tilde{M}=\sum_{i=1}^d\lambda_i\ketbra{\tilde\varphi_i}$
  and $\sum_{i=1}^d\abs{\braket{\tilde\varphi_i}{\varphi_i}}^2=d(1-\er)$, then
  $\norm{\tilde{M}-M}\ge\lambda\sqrt{2\er}$, where
  $\lambda=\min_{i\ne j} \abs{\lambda_i-\lambda_j}$
  and $\norm{\cdot}$ is the Hilbert-Schmidt norm.
\end{obs}

From the definition of the Hilbert-Schmidt norm,
\begin{equation}
  \norm{\tilde{M}-M}^2
  =\Tr(M^2)+\Tr(\tilde{M}^2)-2\Tr(M\tilde{M})
  =2\qty(\sum_{i=1}^d\lambda_i^2
  -\sum_{i,j=1}^dF_{ij}\lambda_i\lambda_j),
\end{equation}
where
\begin{equation}
  F_{ij}=\braket{\tilde\varphi_i}{\varphi_j}
  \braket{\varphi_j}{\tilde\varphi_i}
  =\abs{\braket{\tilde\varphi_i}{\varphi_j}}^2.
\end{equation}
One can easily verify that $F=[F_{ij}]_{i,j=1}^d$ is a doubly stochastic
matrix, and thus can be written as a convex combination of permutation matrices
according to the Brikhoff theorem \cite{BhatiaMatrixanalysis2013}, i.e.,
\begin{equation}
  F=\sum_{\pi\in S_d}p_\pi R_\pi,
  \label{eq:Brikhoff}
\end{equation}
where $S_d$ is the symmetric group of degree $d$,
$R_\pi=\sum_{i=1}^d\ketbra{i}{\pi(i)}$ are the
corresponding permutation matrices, and $p_\pi$ form a probability
distribution. Then,
\begin{equation}
  \norm{\tilde{M}-M}^2
  =2\sum_{\pi\ne e}p_\pi\qty(\sum_{i=1}^d\lambda_i^2
  -\sum_{i=1}^d\lambda_i\lambda_{\pi(i)}).
\end{equation}
where $e$ is the identity element in group $S_d$.
In the following, we will prove that
\begin{equation}
  \sum_{\pi\ne e}p_\pi\ge\varepsilon,
  \label{eq:permutationCoefficients}
\end{equation}
and for any $\pi\ne e$
\begin{equation}
  \sum_{i=1}^d\lambda_i^2
  -\sum_{i=1}^d\lambda_i\lambda_{\pi(i)}
  \ge\lambda^2.
  \label{eq:permutationDiff}
\end{equation}
With Eqs.~(\ref{eq:permutationCoefficients},\,\ref{eq:permutationDiff}), one
can directly show that
$\norm{\tilde{M}-M}\ge\lambda\sqrt{2\varepsilon}$.

Equation~\eqref{eq:permutationCoefficients} follows from the observation that
$1-\Tr(F)/d$ is exactly the measurement infidelity, i.e.,
\begin{equation}
  \Tr(F)=d(1-\varepsilon)
\end{equation}
and by Eq.~\eqref{eq:Brikhoff},
\begin{equation}
  \Tr(F)=
  \sum_{\pi\in S_d}p_\pi\Tr(R_\pi)
  \ge p_e\Tr(\mi_d)=d p_e.
\end{equation}
thus we get that $p_e\le 1-\er$, from which
Eq.~\eqref{eq:permutationCoefficients} follows.

To prove Eq.~\eqref{eq:permutationDiff}, we assume that
$\lambda_1>\lambda_2>\dots>\lambda_d$ without loss of generality.  Then if
$\pi\ne e$, the sequence
$\lambda_{\pi(1)},\lambda_{\pi(2)},\dots,\lambda_{\pi(d)}$ is not sorted.
A well-known sorting algorithm is called the Bubble sort, whose basic step is
to swap the element $\lambda_{\pi(j)}$ and $\lambda_{\pi(j+1)}$ if
$\lambda_{\pi(j)}<\lambda_{\pi(j+1)}$ for some $j$.  Let
$\lambda_{\pi'(1)},\lambda_{\pi'(2)},\dots,\lambda_{\pi'(d)}$ be the sequence
after the swap, then one can easily verify that for each swap of this kind, the
inner product $\sum_{i=1}^d\lambda_i\lambda_{\pi'(i)}$ increases by at least
$\lambda^2$, i.e.,
\begin{equation}
    \sum_{i=1}^d\lambda_i\lambda_{\pi'(i)}-\sum_{i=1}^d\lambda_i\lambda_{\pi(i)}
    =\lambda_j\lambda_{\pi(j+1)}+\lambda_{j+1}\lambda_{\pi(j)}
    -\lambda_j\lambda_{\pi(j)}-\lambda_{j+1}\lambda_{\pi(j+1)}
    =(\lambda_j-\lambda_{j+1})(\lambda_{\pi(j+1)}-\lambda_{\pi(j)})
    \ge\lambda^2,
\end{equation}
where $\lambda=\min_{i\ne j}\abs{\lambda_i-\lambda_j}$. Thus, we prove that for
any $\pi\ne e$, Eq.~\eqref{eq:permutationDiff} holds, as at least one swap is
needed to obtain the sorted sequence $\lambda_1,\lambda_2,\dots,\lambda_d$, for
which the inner product is $\sum_{i=1}^d\lambda_i^2$.

\section{Details of the general results}
\label{appendix:obs}

\textit{Proof of Eq.~\eqref{eq:boundGeneralMbar}.}---%
Equation~\eqref{eq:Mbar-d} implies that, for the target
measurement $\qty{P_i=\ketbra{\varphi_i}}_{i=1}^d$, the corresponding tuned
measurement $ \qty{\bar{M}_i}_{i=1}^d$ is diagonal in the basis
$\qty{\ketbra{\varphi_i}}_{i=1}^d$. Thus, each $\bar{M}_i$ admits the form
$\bar{M}_i=\sum_{k=1}^da_k\ketbra{\varphi_k}$, where $0\le a_k\le 1$. Since
\begin{equation}
  \Tr(P_i\bar{M}_i) + \Tr[(\mi_d- P_i)(\mi_d-\bar{M}_i)]\ge \sum_{k=1}^d
  \Tr(P_k\bar{M}_k)
\end{equation}
and
\begin{equation}
  \er=1-\frac{1}{d}\sum_{k=1}^d \Tr(P_k\bar{M}_k),
\end{equation}
it holds that
\begin{equation}
  a_i - \sum_{k\ne i} a_k \ge 1 - d\er.\label{eq:coefficient}
\end{equation}
Taking the constraints $0\le a_k\le 1$ into consideration, we obtain that
\begin{equation}
  a_i\ge 1 - d\er, \qand a_k\le d\er \qfor k\ne i,
\end{equation}
which further imply that
\begin{equation}
  (1 - d\er)P_i \le \bar M_i \le (1 - d\er)P_i + d\er \mi_d.
\end{equation}

\textit{Proof of Eq.~\eqref{eq:generalbound}.}---%
In the proof, we assume that $d_n\er<1$ for all $n$, because what we concern is
the scaling of the witness error when $\er\to0$. If two sets of positive
semidefinite operators $\{A_n\}$ and $\{B_n\}$ satisfy that $A_n\ge B_n$ for
each $n$, then it holds that $\bigotimes_n A_n\ge\bigotimes_n B_n$.
Thus, Eqs.~(\ref{eq:Wtuned},\,\ref{eq:boundGeneralMbar}) imply that
\begin{equation}
  \bar W_\er\ge W^+ +W^-,
\end{equation}
where
\begin{equation}
W^+=\sum_{w_\mu>0}w_\mu\bigotimes_{n=1}^N\qty(1-d_n\er)\ope{P}{\mu}{n}\qand
W^-=\sum_{w_\mu<0}
w_\mu\bigotimes_{n=1}^N\qty[\qty(1-d_n\er)\ope{P}{\mu}{n}+d_n\er\mi_{d_n}].
\end{equation}
Moreover, $W^-$ can be decomposed into
\begin{equation}
  W^-=W_0^-+\er W_1^-+\er^2 W_2^-+\dots +\er^N W_N^-,
\end{equation}
where
\begin{equation}
  W_0^-=\sum_{w_\mu<0}w_\mu\bigotimes_{n=1}^N\qty(1-d_n\er)
  \ope{P}{\mu}{n}\qc
  W_1^-=\sum_{w_\mu<0}w_\mu\sum_{n=1}^N\bigotimes_{n'=1}^N
  \ope{\mathcal{P}}{\mu}{n,n'},
\end{equation}
$\ope{\mathcal{P}}{\mu}{n,n'}=d_n\mi_{d_n}$ if $n=n'$ and
$\ope{\mathcal{P}}{\mu}{n,n'}=\ope{P}{\mu}{n}$ otherwise,
and $W_2^-,\dots,W_N^-$ are some Hermitian operators independent of $\er$.
Note that
\begin{equation}
  W^+ + W_0^-=\qty[\prod_{n=1}^N\qty(1-d_n\er)]W,
\end{equation}
then it holds that
\begin{equation}
  \bar W_\er\ge \qty[\prod_{n=1}^N\qty(1-d_n\er)]W+\er W_1^-+\er^2
  W_2^-+\dots+\er^N W_N^-.
\end{equation}
Thus, for all states that cannot be detected by $W$, i.e., those states such
that $\expval{W}\ge0$, we have
\begin{equation}
  \expval{\bar W_\er}\ge \er \expval{W_1^-} + O(\er^2).
\end{equation}
Since $\ope{P}{\mu}{n}\le\mi_{d_n}$, we have
$\bigotimes_{n'=1}^N\ope{\mathcal{P}}{\mu}{n,n'}\le d_{n}\mi_{d_1d_2\dots
d_N}$, then it follows that
\begin{equation}
  \expval{\bar W_\er}\ge\er\sum_{w_\mu<0}w_\mu\sum_{n=1}^N d_{n}+O\qty(\er^2).
  \label{eq:generalboundA}
\end{equation}

\begin{figure}
  \includegraphics[width=.6\linewidth]{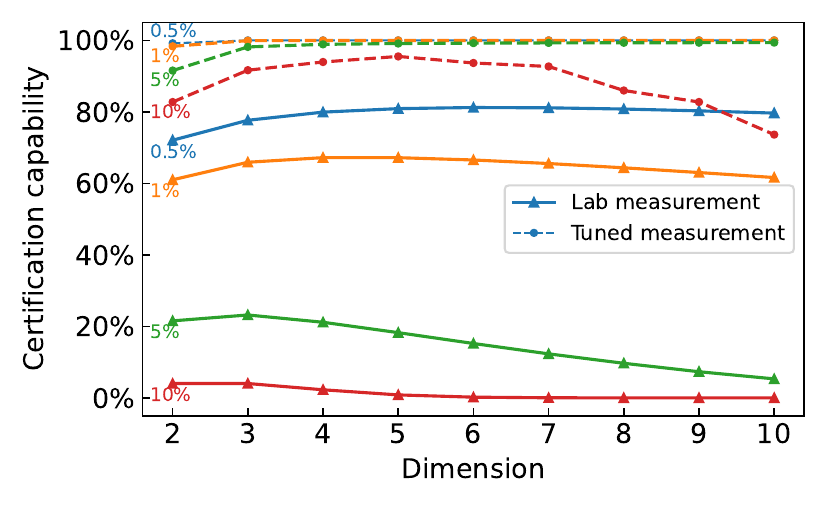}
  \caption{The comparison of certification capability between the witnesses
    $W=\frac{d+1}{d}\mi_{d^2} - \sum_{k=1}^d (\ketbra{e_k,e_k} +
    \ketbra{f_k,f_k^*})$,
    with the laboratory measurements and the tuned measurements for systems
    with different dimensions ($2,3,\dots,10$) under different infidelities
    ($0.5\%,1\%,5\%,10\%$). Triangle symbols
    and solid lines represent the certification capability with the lab
    measurements, while dot symbols and dashed lines represent those with the
    tuned measurements. The certification capability of the witnesses improves
    remarkably with tuned measurements.
  }
  \label{fig:comparison2}
\end{figure}

\textit{Examples.}---%
Equation~\eqref{eq:generalboundA} implies that randomized measurements can
significantly suppress the measurement errors in the general case.
In addition, we would like to mention that the bound
in Eq.~\eqref{eq:generalboundA} is usually not optimal, hence the efficiency of our
randomized-measurement-based method can be further improved.
In contrary to Eq.~\eqref{eq:generalboundA}, the exact bound
$\mathcal{B}^{\mathrm{rand}}(\er)$ does not significantly depend on the
dimension for many entanglement witnesses.
In Fig.~\ref{fig:comparison2}, we illustrate this with the entanglement
witnesses based on mutually unbiased bases
\cite{SpenglerEntanglementdetectionvia2012}.
We consider the following witness in a two-qudit system
\begin{equation}
  W^{(d)} = \frac{d+1}{d}\mi_{d^2} - \sum_{k=1}^d \qty(\ketbra{e_k,e_k}
  + \ketbra{f_k,f_k^*}),
\end{equation}
where $\qty{\ket{e_k}}_{k=1}^d$ is an orthonormal basis and
$\qty{\ket{f_k}}_{k=1}^d$ is the Fourier basis, i.e., $\ket{f_k}
= \frac{1}{\sqrt{d}}\sum_{\ell=1}^d \ee^{\frac{2\pi\ii}{d}
k\ell}\ket{e_{\ell}}$ and $\ket{f_k^*} = \frac{1}{\sqrt{d}}\sum_{\ell=1}^d
\ee^{-\frac{2\pi\ii}{d}k\ell}\ket{e_{\ell}}$.
The original certification range is $[-\frac{d-1}{d},0)$, and
the certification capability under laboratory measurements and tuned measurements
are $1+\frac{d}{d-1}\mathcal{B}(\er)$ and
$1+\frac{d}{d-1}\mathcal{B}^\mathrm{rand}(\er)$, respectively.
We have used the alternating convex search method to
numerically calculate the certification capability of the witnesses for systems
with different dimensions under different measurement infidelities. The
numerical results suggest that our method suppresses the errors effectively.
From Fig.~\ref{fig:comparison2}, one can see that before applying our method,
the certification capability is severely compromised, even if the measurement
infidelity $\er$ is relatively small, while after applying our method almost
all certification capability is restored.
Even when the measurement infidelity reaches a relatively large value of $10\%$
for which the certification capability of the laboratory measurements almost vanishes,
our method can still effectively suppress the errors and restore over $70\%$ of
the certification capability. Moreover, in this example, the optimal bound does
not significantly depend on the dimension.


\textit{Implementation with discrete randomized measurements.}---%
In passing, we would like to mention that the implementation of randomized
measurements can be also discretized, making it more feasible for some
quantum systems.
We define $G$ as the group
generated by the set of unitary operators $\qty{U_k}_{k=1}^d$, where
$U_k\ket{\varphi_{\ell}}=(-1)^{\delta_{k\ell}}\ket{\varphi_{\ell}}$, i.e.,
$U_k\ket{\varphi_{\ell}}=-\ket{\varphi_{\ell}}$ when $k=\ell$ and
$U_k\ket{\varphi_{\ell}}=\ket{\varphi_{\ell}}$ otherwise. For any POVM
element $M$, the average under group $G$,
$\bar{M}:=\frac{1}{\abs{G}}\sum_{U\in G}U^\dagger MU
=\sum_{k=1}^d\expval{M}{\varphi_k}\ketbra{\varphi_k}$,
where $\abs{G}$ is the number of elements in group $G$.
This attains the same result as Eq.~\eqref{eq:Mbar-d}, and thus $G$ can be
adopted for discretizing the randomized measurements. The elimination of the
off-diagonal terms in this case follows from the simple facts
that $U_k^\dagger\bar{M}U_k=\bar{M}$ for any $k$, and
$\mel{\varphi_k}{U_k^\dagger\bar{M}U_k}{\varphi_\ell}
=-\mel{\varphi_k}{\bar{M}}{\varphi_\ell}$
when $k\ne\ell$.

\textit{Alternative implementation.}---%
In the main text we use the randomized unitary operation $U_{\boldsymbol{\theta}} = \sum_{k=1}^d \ee^{\ii\theta_k} \ketbra{\varphi_k}$ to suppress errors, since $\mathbb{E}_{\boldsymbol{\theta}} (U_{\boldsymbol{\theta}}^\dagger M_i U_{\boldsymbol{\theta}}) = \sum_{k=1}^d \expval{M_i}{\varphi_k} \ketbra{\varphi_k}$. Alternatively, the randomized unitary operation $U_\theta' = \sum_{k=1}^d \ee^{\ii(k-1)\theta} \ketbra{\varphi_k}$, with $\theta$ chosen uniformly from $[0, 2\pi)$, also satisfies this condition, since $U_\theta^{\prime\dagger} M_i U_\theta' = \sum_{k, l=1}^d \ee^{-\ii(k-l)\theta} \mel{\varphi_k}{M_i}{\varphi_l} \ketbra{\varphi_k}{\varphi_l}$ and $\mathbb{E}_\theta[\ee^{-\ii(k-l)\theta}] = \delta_{kl}$.

Moreover, this implementation can also be discretized. The set of unitary
operators is $\{U_j = \sum_{k=1}^d \ee^{(k-1)j
\frac{2\pi\ii}{d}}\ketbra{\varphi_k}\}_{j=1}^d$. Then we have $U_j^\dagger M_i
U_j = \sum_{k, l=1}^d \ee^{-(k-l)j \frac{2\pi\ii}{d}}
\mel{\varphi_k}{M_i}{\varphi_l} \ketbra{\varphi_k}{\varphi_l}$ and $\frac{1}{d}
\sum_{j=1}^d \ee^{-(k-l)j \frac{2\pi\ii}{d}} = \delta_{kl}$, which can be used
to suppress measurement errors.

\section{Gate-independent errors}\label{appendix:gate_independent}

Recall that our randomized-measurement-based method is to additionally
perform randomized unitary operations $U_{\vect{\theta}}$ on the quantum state
$\rho$ before the laboratory measurement $\{M_i\}_{i=1}^d$.
Now suppose that the randomized operations are affected by gate-independent
errors, that is, the state after the imprecise randomized operation is
described by $\mathcal{E}(U_{\vect{\theta}} \rho U_{\vect{\theta}}^\dagger)$.
Here, $\mathcal{E}$ denotes the noisy channel, which is a completely positive
and trace-preserving (CPTP) map independent of $\vect{\theta}$.
As $\Tr[\mathcal{E} (U_{\vect{\theta}} \rho U_{\vect{\theta}}^\dagger) M_i]
= \Tr[U_{\vect{\theta}} \rho U_{\vect{\theta}}^\dagger \mathcal{E}^*(M_i)]$,
the imprecise randomized operations before the laboratory measurement
$\{M_i\}_{i=1}^d$ are equivalent to perfect randomized operations before the
measurement $\{\mathcal{E}^*(M_i)\}_{i=1}^d$, where $\mathcal{E}^*(\cdot)$ is
the dual map of $\mathcal{E}(\cdot)$, that is,
$\mathcal{E}^*(\cdot)$ satisfies that
$\Tr[\mathcal{E}(X)Y]=\Tr[X\mathcal{E}^*(Y)]$.
When the errors are small, the infidelity of the measurement
$\{\mathcal{E}^*(M_i)\}_{i=1}^d$ should also be small.
Quantitatively, if the minimum gate
fidelity~\cite{GilchristDistancemeasurescompare2005} of $\mathcal{E}$ and the
identity map $\mathcal{I}$ is $1-\tau$ and the infidelity of the lab
measurement $\{M_i\}_{i=1}^d$ is $\er$, then the infidelity of the measurement
$\{\mathcal{E}^*(M_i)\}_{i=1}^d$ is no larger than
$(\sqrt{\er}+\sqrt{\tau})^2$; see the observation below.
Typically, the precision of the quantum gates is one or two orders of
magnitude better than the measurements in the same physical system.
This validates our randomized-measurement-based method in the presence of
gate-independent errors.
Moreover, the analysis also demonstrates that our method
remains effective even when the precision of the quantum gates is comparable to
that of the measurements.

\begin{obs}
If the minimum gate fidelity of $\mathcal{E}$ with respect to the
identity map $\mathcal{I}$ is $1-\tau$, and the infidelity of the lab
measurement $\{M_i\}_{i=1}^d$ with respect to the target measurement
$\{P_i=\ketbra{\varphi_i}\}_{i=1}^d$ is $\er$, then the infidelity of the
measurement $\{\mathcal{E}^*(M_i)\}_{i=1}^d$ with respect to the target
measurement is no larger than $(\sqrt{\er}+\sqrt{\tau})^2$.
\end{obs}

Note that the minimal fidelity $1-\tau$ of $\mathcal{E}$ and identity map
$\mathcal{I}$ is defined as \cite{GilchristDistancemeasurescompare2005}
\begin{equation}
	1-\tau=F_{\min}(\mathcal{E},\mathcal{I}):=\min_{\ket{\psi}}
	F(\ket{\psi},\mathcal{E}(\ketbra{\psi})),
\end{equation}
where $F(\ket{\psi},\sigma)=\expval{\sigma}{\psi}$ denotes
the fidelity of two states $\ket{\psi}$ and $\sigma$,
the infidelity $\er$ of the lab
measurement $\{M_i\}_{i=1}^d$ with respect the target measurement
$\{P_i=\ketbra{\varphi_i}\}_{i=1}^d$ is defined as
\begin{equation}
  \er=1-\frac{1}{d}\sum_{i=1}^d\Tr(P_iM_i)
  =1-\frac{1}{d}\sum_{i=1}^d\expval{M_i}{\varphi_i},
  \label{eq:infidelityM}
\end{equation}
and we aim to prove that the infidelity $\mu$ of the measurement
$\{\mathcal{E}^*(M_i)\}_{i=1}^d$ with respect to
$\{P_i=\ketbra{\varphi_i}\}_{i=1}^d$ is no larger than
$(\sqrt{\er}+\sqrt{\tau})^2$, i.e.,
\begin{equation}
  \mu=1-\frac{1}{d}\sum_{i=1}^d\Tr(P_i\mathcal{E}^*(M_i))
  =1-\frac{1}{d}\sum_{i=1}^d\Tr(\mathcal{E}(P_i)M_i)
  \le(\sqrt{\er}+\sqrt{\tau})^2
  \label{eq:infidelityEM}
\end{equation}

Let us denote $\mathcal{E}(P_i)$ as $\rho_i$, and
$\Tr(P_iM_i)=\expval{M_i}{\varphi_i}$ as $1-\er_i$, then
$\expval{\rho_i}{\varphi_i}\ge 1-\tau$, $\er_i\ge 0$, and
$\frac{1}{d}\sum_{i=1}^d\er_i=\er$ due to Eq.~\eqref{eq:infidelityM}.
The bound in Eq.~\eqref{eq:infidelityEM} will result from the upper and lower
bounds of $\frac{1}{d}\sum_{i=1}^d
\Tr[(\ketbra{\varphi_i}-\rho_i)(M_i-\ketbra{\varphi_i})]$, i.e.,
\begin{equation}
  \mu-\er-\tau\le\frac{1}{d}\sum_{i=1}^d
  \Tr[(\ketbra{\varphi_i}-\rho_i)(M_i-\ketbra{\varphi_i})]
  \le2\sqrt{\er\tau}.
\end{equation}

To obtain the lower bound, we perform a direct calculation,
\begin{equation}
  \frac{1}{d}\sum_{i=1}^d\Tr[(\ketbra{\varphi_i}-\rho_i)
  (M_i-\ketbra{\varphi_i})]
  =-\frac{1}{d}\sum_{i=1}^d\Tr(\rho_iM_i)
  +\frac{1}{d}\sum_{i=1}^d\expval{M_i}{\varphi_i}
  +\frac{1}{d}\sum_{i=1}^d\expval{\rho_i}{\varphi_i}
  -1,
\end{equation}
then the relations
\begin{align}
  \mu&=1-\frac{1}{d}\sum_{i=1}^d\Tr(\rho_iM_i),\\
  \er&=1-\frac{1}{d}\sum_{i=1}^d\expval{M_i}{\varphi_i},\\
  \tau&\ge 1-\expval{\rho_i}{\varphi_i},
\end{align}
imply that
\begin{equation}
  \frac{1}{d}\sum_{i=1}^d
  \Tr[(\ketbra{\varphi_i}-\rho_i)(M_i-\ketbra{\varphi_i})]
  \ge\mu-\er-\tau.
\end{equation}

To obtain the upper bound, we take advantage of the Cauchy-Schwarz inequality,
\begin{equation}
  \Tr[(\ketbra{\varphi_i}-\rho_i)(M_i-\ketbra{\varphi_i})]
  \le\norm{\ketbra{\varphi_i}-\rho_i}\norm{M_i-\ketbra{\varphi_i}},
  \label{eq:boundBeforeRelax}
\end{equation}
where $\norm{\cdot}$ is the Hilbert-Schmidt norm.
From the definition of the Hilbert-Schmidt norm, we get that
\begin{equation}
  \norm{\ketbra{\varphi_i}-\rho_i}^2
  =\Tr(\rho_i^2)+1-2\expval{\rho_i}{\varphi_i}
  \le 2\tau,
\end{equation}
and
\begin{equation}
  \norm{M_i-\ketbra{\varphi_i}}^2
  =\Tr(M_i^2)+1-2\expval{M_i}{\varphi_i}
  \le\Tr(M_i)-1+2\er_i,
\end{equation}
where we have used the relations that $\Tr(\rho_i^2)\le 1$ and
$\Tr(M_i^2)\le\Tr(M_i)$. Thus,
\begin{equation}
  \begin{aligned}
    &\frac{1}{d}\sum_{i=1}^d
    \Tr[(\ketbra{\varphi_i}-\rho_i)(M_i-\ketbra{\varphi_i})]\\
    \le&\frac{1}{d}\sum_{i=1}^d
    \norm{\ketbra{\varphi_i}-\rho_i}\norm{M_i-\ketbra{\varphi_i}}\\
    \le&\frac{1}{d}\sum_{i=1}^d\sqrt{2\tau}\sqrt{\Tr(M_i)-1+2\er_i}\\
    \le&\sqrt{2\tau}\sqrt{\frac{1}{d}\sum_{i=1}^d\qty[\Tr(M_i)-1+2\er_i]}\\
    =&2\sqrt{\er\tau},
  \end{aligned}
\end{equation}
where we have used the concavity of the square root function for the last
inequality, and the relations that $\sum_{i=1}^dM_i=\mi_d$ and
$\frac{1}{d}\sum_{i=1}^d\er_i=\er$ for the last equality.

\section{Gate-dependent errors}\label{appendix:gate_dependent}

\begin{figure}
  \centering
  \includegraphics[width=.6\linewidth]{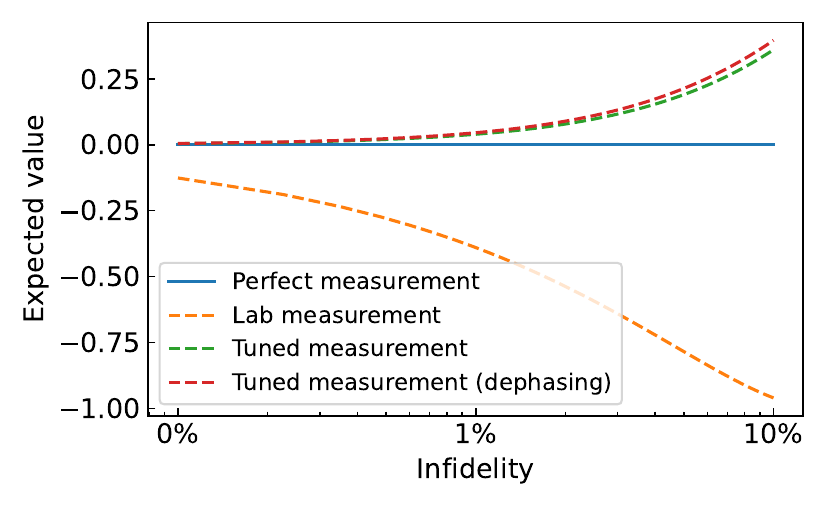}
  \caption{The expected values of the witness
    $\mi_4-\sx^A\ot\sx^B-\sz^A\ot\sz^B$, with the imprecise measurements in
  Eqs.~(\ref{eq:lab_x},\,\ref{eq:lab_z}).
    Note that, in this case, the tuned measurements do not saturate the
    separable bound in Eq.~\eqref{eq:ModifiedBound} in contrast to
    Fig.~\ref{fig:comparison} in the main text.
  }
  \label{fig:dephasing_misalignment}
\end{figure}

\begin{figure}
    \centering
    \includegraphics[width=\linewidth]{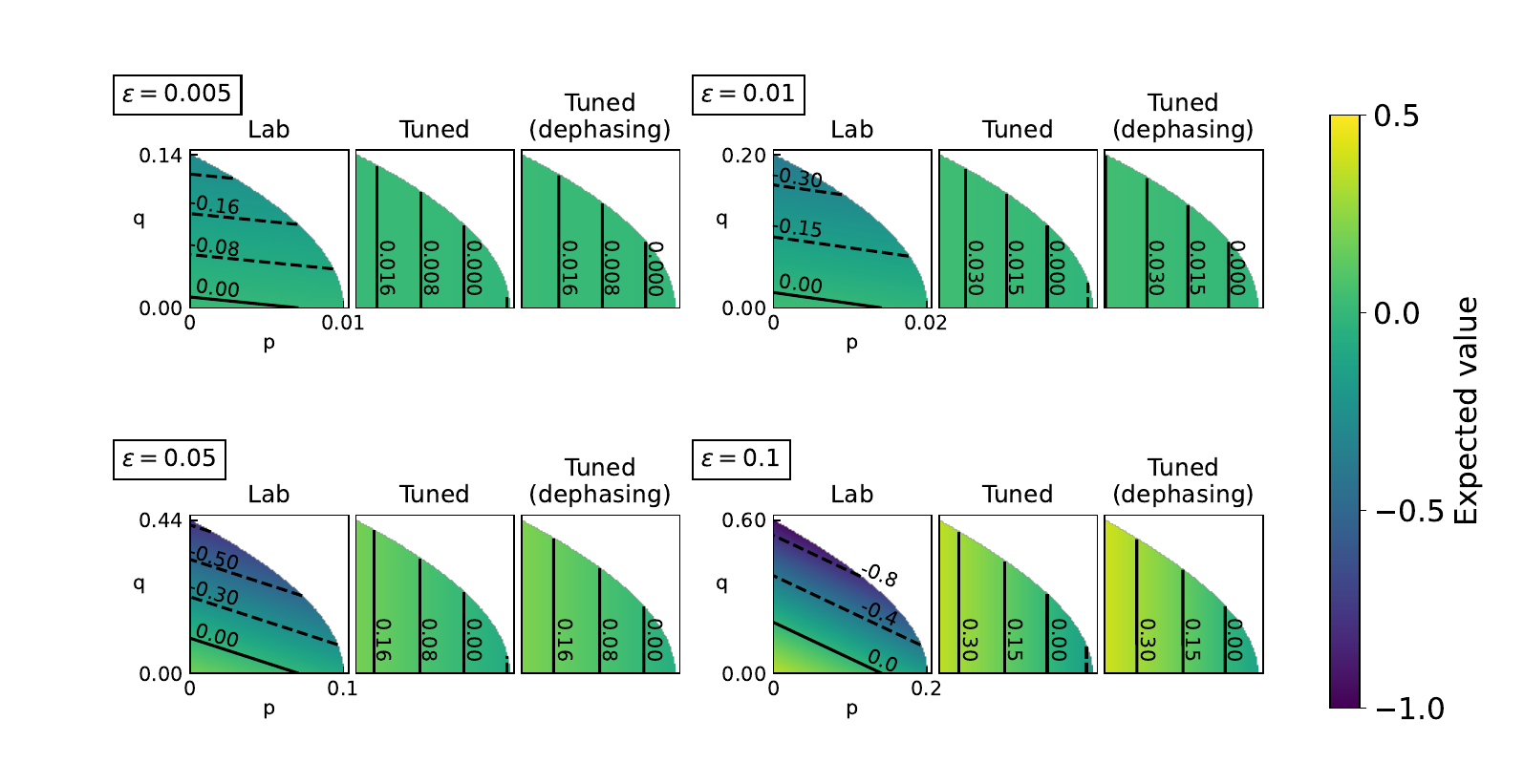}
    \caption{The expected values of the witness
      $\mi_4-\sx^A\ot\sx^B-\sz^A\ot\sz^B$
      with general imprecise measurements of different infidelities.
      $p$ and $q$ are the parameters in
      Eqs.~(\ref{eq:pq_measurementx},\,\ref{eq:pq_measurementz}), which denote
      the magnitude of the commutative terms and the noncommutative terms of
      the laboratory measurements, respectively.
    }
    \label{fig:dephasing_general}
\end{figure}

To investigate the impact of gate-dependent errors, we examine the most
common noise in experiments: dephasing. The results show that our
method maintains its effectiveness when the randomized operations are subject
to dephasing. Still, we consider the witness $W=\mi_4 - \sx^A\otimes\sx^B
- \sz^A\otimes\sz^B$ in a two-qubit system. The system is in state
$\rho=\ketbra{\varphi}\otimes\ketbra{\varphi}$ with
$\ket{\varphi}=\cos(\pi/8)\ket{0}+\sin(\pi/8)\ket{1}$,
which satisfies that $\Tr(W\rho)=0$ with perfect measurements,
while the actual measurements are always subject to errors. In the main text,
we have shown that performing randomized
measurements can significantly suppress the measurement errors.
While the randomized operations can also be affected by noise, the actual gate
will be described by the CPTP map $\mathcal{E}_\theta$ instead of
$U_\theta=\diag(1,\ee^{\ii \theta})$. For dephasing noise, the actual
gate can be derived from the following Lindblad master equation:
\begin{equation}
  \frac{\dd{\rho(t)}}{\dd{t}} = - \frac{i}{\hbar}[H, \rho(t)] + \gamma \qty[L
    \rho(t) L^\dagger - \frac{1}{2}\{L^\dagger L, \rho(t)\}],
\end{equation}
where $H$ is the Hamiltonian of the system, and $L=\sqrt{\gamma/2}\sz$ is the
Lindblad operator representing decoherence process that occurs at rate
$\gamma$. To implement the gate $U=\diag(1,\ee^{\ii
\theta})$ under the measurement bases of $\sz$ and $\sx$, we take $H=\hbar
\omega \sz/2$ and $H=\hbar \omega \sx/2$, respectively, where $\hbar \omega$
denotes the energy level difference. Typically, the precision of the quantum
gate is one or two orders of magnitude better than the measurements in
the same physical systems. Therefore, in the simulation, we choose the mean
value of the average gate fidelity~\cite{GilchristDistancemeasurescompare2005}
$F^\mathrm{ave}(\mathcal{E}_\theta, U_\theta)$ over $\theta\in [0,2\pi)$
to be
\begin{equation}
  \mathbb{E}_{\theta}
  [F^\mathrm{ave}(\mathcal{E}_\theta, U_\theta)]
  =\frac{1}{2\pi}\int_{0}^{2\pi}
  F^\mathrm{ave}(\mathcal{E}_\theta, U_\theta) \dd\theta
  =1-\frac{\er}{10}.
\end{equation}

In Fig.~\ref{fig:dephasing_misalignment}, the imprecise measurements are taken as
\begin{align}
    M_x^A = M_x^B = (1-2\er)\sx + 2\sqrt{\er(1-\er)}\sz,\label{eq:lab_x}\\
    M_z^A = M_z^B = (1-2\er)\sz + 2\sqrt{\er(1-\er)}\sx,\label{eq:lab_z}
\end{align}
where $\er$ is the infidelity of the measurements.
  Then we apply our method to these measurements.
The results demonstrate the
advantage of our method in handling misalignment errors.
Similar results also hold for general measurement errors besides the misalignment.
In Fig.~\ref{fig:dephasing_general}, we consider
\begin{align}
    M_x^A=M_x^B=p\mi_2 + (1-2\er)\sx
    + q\sz,\label{eq:pq_measurementx}\\
    M_z^A=M_z^B=p\mi_2 + (1-2\er)\sz
    + q\sx,\label{eq:pq_measurementz}
\end{align}
where $p$ denotes the magnitude of commutative
terms, $q$ denotes the magnitude of noncommutative terms,
and $|p|+\sqrt{(1-2\er)^2+q^2}\le1$.
As the results demonstrated, our method consistently outperforms lab
measurements and shows robustness against dephasing errors.


\section{Entanglement visibility}\label{appendix:visibility}

To further demonstrate the performance of our method, we investigate
entanglement visibility for the witness $W=\mi_4 - \sx^A\otimes\sx^B -
\sz^A\otimes\sz^B$. Suppose that the state of system is
$\rho_v = v \ketbra{\phi^+} + \frac{1-v}{4} \mi_4$, where $v \in [0, 1]$ is
the visibility. In the case that the infidelities of local measurements are at
most $\er$, the actual measurements are $M_x^A$, $M_z^A$, $M_x^B$, $M_z^B$
rather than $\sx^A$, $\sz^A$, $\sx^B$, $\sz^B$. To certify the existence of
entanglement reliably, we must have
\begin{align}
  &\Tr[(\mi - M_x \otimes M_x - M_z \otimes M_z)\rho] < \mathcal{B}(\er),\\
  &\Tr[(\mi - \bar{M}_x \otimes \bar{M}_x - \bar{M}_z \otimes \bar{M}_z)\rho] < \mathcal{B}^\mathrm{rand} (\er)
\end{align}
for laboratory measurements and tuned measurements, respectively.

\begin{figure}
  \centering
  \begin{minipage}{.49\textwidth}
    \includegraphics[width=\linewidth]{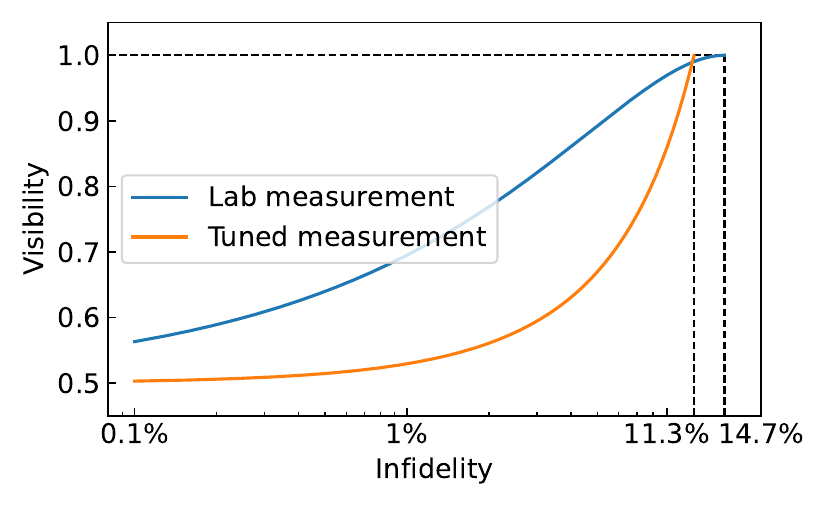}
  \end{minipage}
  \begin{minipage}{.49\textwidth}
    \includegraphics[width=\linewidth]{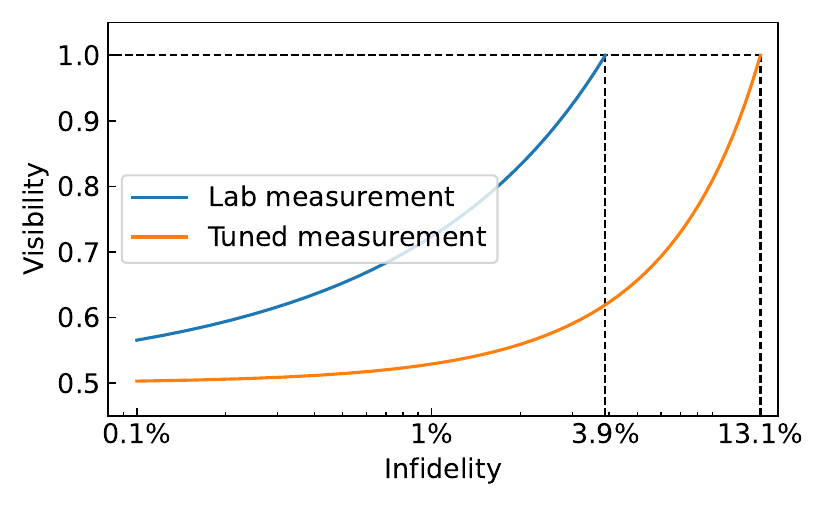}
  \end{minipage}
  \caption{The lower bounds of visibility $v$ of the certifiable state $\rho=v\ketbra{\psi} + \frac{1-v}{4}\mi$. The left figure corresponds to the measurements in Eq.~\eqref{eq:Imp_eg1}, and the right figure corresponds to the measurements in Eq.~\eqref{eq:Imp_eg2}.}
  \label{fig:visibility}
\end{figure}

Obviously, different measurements will lead to different results. 
Here, we consider two sets of imprecise measurements as examples. The first set is the imprecise measurements defined as in Eq.~\eqref{eq:ImpSaturation}, i.e.,
\begin{equation}\label{eq:Imp_eg1}
  M_x^A = M_x^B = (1-2\er)\sx + 2\sqrt{\er(1-\er)}\sz, \quad
  M_z^A = M_z^B = (1-2\er)\sz + 2\sqrt{\er(1-\er)}\sx,
\end{equation}
and the second set is the imprecise measurements defined as in Eq.~\eqref{eq:ModifiedSaturation}, i.e.,
\begin{equation}\label{eq:Imp_eg2}
  M_x^A = M_x^B = 2\er\mi_2 + (1-2\er)\sx, \quad
  M_z^A = M_z^B = 2\er\mi_2 + (1-2\er)\sz.
\end{equation}
The results in Fig.~\ref{fig:visibility} show that our method improves the
certification tasks in general, while under rare circumstances the lab
measurements outperform the tuned measurements.
These exceptions are also interesting because they suggest that errors may be
helpful under certain circumstances.



\twocolumngrid

\bibliography{ref.bib}

\end{document}